\documentclass[11pt]{article}
\usepackage{amsmath}
\usepackage{amsfonts}
\usepackage{amssymb}
\usepackage{theorem}
\newtheorem{theorem}{Proposition}
\newtheorem{lemma}{Lemma}
\newcommand{\dbar}{\bar{\partial}}
\newcommand{\wt}{\widetilde}
\newcommand{\be}{\begin{equation}}
\newcommand{\ee}{\end{equation}}
\newcommand{\bea}{\begin{eqnarray}}
\newcommand{\eea}{\end{eqnarray}}
\newcommand{\beaa}{\begin{eqnarray*}}
\newcommand{\eeaa}{\end{eqnarray*}}

\newcommand{\nn}{\nonumber}
\begin{document}
%%%%%%%%%%%%%%%%%%%%%%%%%%%%%%%%%%%%%%%%%%%%%%%%%%%%%%%%%%%%
%%%%%%%%%%%%%%%%%%%%%%%%%%%%%%%%%%%%%%%%%%%%%%%%%%%%%%%%%%%%%%%%%%%%%%
\title
{Symmetry constraints for dispersionless integrable equations
and systems of hydrodynamic type}
%%%%%%%%%%%%%%%%%%%%%%%%%%%%%%%%%%%%%%%%%%%%%%%%%%%%%%%%%%%%%%%%%%%%%%%%%
%article
%%%%%%%%%%%%%%%%%%%%%%%%%%%%%%%%%%%%%%%%%%%%%%%%%%%%%%%%%%%%%%%%%%%%%%%%
\author{
L.V. Bogdanov\thanks
{L.D. Landau ITP, Kosygin str. 2,
Moscow 119334, Russia} 
~and B.G. Konopelchenko\thanks
{Dipartimento di Fisica dell' Universit\`a di Lecce
and Sezione INFN, 73100 Lecce, Italy}}
\date{}
\maketitle
%%%%%%%%%%%%%%%%%%%%%%%%%%%%%%%%%%%%%%%%%%%%%%%%%%%%%%%%%%%%%%%%%%%%%%
\begin{abstract}
Symmetry constraints for (2+1)-dimensional dispersionless
integrable equations  are considered. It is demonstrated that they naturally lead to systems 
of hydrodynamic type which arise within the reduction method.
One also easily obtaines an associated complex curve
(Sato function) and corresponding
generating equations. Dispersionless KP and 2DTL hierarchy 
are considered as illustrative examples.
\end{abstract}
%%%%%%%%%%%%%%%%%%%%%%%%%%%%%%%%%%%%%%%%%%%%%%%%%%%%%%%%%%%%%%%%%
\section{Introduction}
Dispersionless integrable equations arise in various contexts
and have variety of applications from complex analysis
to topological field theory 
(see \cite{KM77}-\cite{DMT}).
The reduction method seems to be the most developed method of
construction of exact solutions of dispersionless equations 
\cite{4,YK,TT,16},\cite{FKS}-\cite{CheTu}.
It reveals a close connection between such equations and certain systems
of hydrodynamic type, emphasising a role of complex curves
(Sato functions).
The reduction method has been successfully applied to
several dispersionless equations. However, an intrinsic
background of its validity and efficiency seems to remain not very clear.

In the present paper it is shown that systems of hydrodynamic type
arising within the reduction method are nothing but the symmetry constrained (2+1)-dimensional integrable equations. We will consider
here the dispersionless KP and 2DTL hierarchies.

First, we find a general infinitesimal `nonisospectral'
symmetry transformation of the dispersionless
$\tau$-function (free energy $F$). Then we impose symmetry constraint
of the type `isospectral symmetry = nonisospectral symmetry',
similar to dispersionfull case \cite{KStr}-\cite{Orlov91}.
It is demonstrated that such a constraint reduces (2+1)-dimensional
dispersionless equation to a set of (1+1)-dimensional systems of
hydrodynamic type with a finite number of dependent
variables. We present also a simple way to find an associated
complex curve and to construct generating equations.

Our approach starts with the $\dbar$-method applied to
dispersionless equations. It provides us with a simple and natural
way to consider symmetries and symmetry constraints. At the same time,
an analysis of such constraints indicates a direction of a
possible extension of the
quasiclassical $\dbar$-method.
%%%%%%%%%%%%%%%%%%%%%%%%%%%%%%%%%%%%%%%%%%%%%%%%%%%%%%%%%%%%%%%%%%
%\section{Quasiclassical $\dbar$-method,
%$\tau$-function and symmetries}
%%%%%%%%%%%%%%%%%%%%%%%%%%%%%%%%%%%%%%%%%%%%%%%%%%%%%%%%%%%%%%%%%
\section{Quasiclassical $\dbar$-method}
Dispersionless integrable hierarchies can be described in different
forms within different approaches. In the papers \cite{KMR,KM1,KM2}
it was shown that such hierarchies can be introduced starting with the 
nonlinear Beltrami equation (quasi-classical $\dbar$-problem)
\bea
%\begin{equation}
S_{\bar z}=W(z, \bar z, S_z),
\label{ddbar}
%\end{equation}
\eea
where $z\in\mathbb{C}$, bar means complex conjugation,
$
S_z=\frac{\partial S(z,\bar{z})}{\partial z},
$
$
S_{\bar z}=\frac{\partial S(z,\bar{z})}{\partial \bar z},
$
and $W$ (quasi-classical $\dbar$-data) is an analytic function
of $S_z$,
\bea
&&
W(z,\bar z,S_z)=\sum_{p=0}^{\infty}w_p(z,\bar z)(S_z)^p.
\label{Wform}
\eea
Applying the quasi-classical $\dbar$-dressing method
based on equation (\ref{ddbar}), one can get dispersionless
integrable hierarchies and the corresponding addition formulae
in a very regular and simple way. Such an approach reveals also
the connection of dispersionless hierarchies with the quasi-conformal
mappings on the plane.

The specific choice of the hierarchy depends on the choice of
domain $G$ which is the support of the $\dbar$-data. Once
the domain is fixed, we introduce the class of functions $S_0(z)$,
having the derivative $\partial_z S_0(z)$ analytic in $G$,
and formulate the boundary problem in $G$ in the following way.
Let the function $S_0(z)$ be given.
The problem is to find the function $S=S_0+\wt S$, satisfying
(\ref{ddbar}) in $G$, with $\wt S$ analytic outside $G$
and decreasing at infinity 

Introducing parameterization of the class of functions $S_0(z)$ in 
terms of infinite number of variables (times), and using
the technique of quasi-classical $\dbar$-dressing method, 
it is possible to demonstrate that $S(z,\bar z,\mathbf{t})$ is a 
solution
of Hamilton-Jacobi equations
of corresponding dispersionless hierarchy.

For dispersionless KP (dKP) hierarchy we take unit disc $D$ as $G$, and parameterization
of $S_0(z)$ is given by
\be
S_0(z,\mathbf{t})=\sum_{n=1}^\infty t_n z^n,
\label{paramKP}
\ee
$S(z,\bar z,\mathbf{t})=S_0+\wt S$,
$\wt S$ is analytic outside the unit disc, and
at infinity it has an expansion
$\wt S=\sum_{i=1}^{\infty}S_i(\mathbf{t})z^{-i}$. 
The quantity $p=\frac{\partial S}{\partial t_1}$
is a basic homeomorphism \cite{KM1}.
Important role in the theory of dKP hierarchy
is played by the equation
\be
p(z)-p(z_1)+z_1\exp(-D(z_1)S(z))=0,\quad z\in\mathbb{C},
\quad z_1\in\mathbb{C}\setminus D,
\label{DE2}
\ee
(where $D(z)$ is the quasiclassical vertex operator,
$D(z)=\sum_{n=1}^{\infty}\frac{1}{n}\frac{1}{z^n}
\frac{\partial}{\partial t_n}$, $|z| > 1$) 
which generates Hamilton-Jacobi equations of the hierarchy
by expansion into the powers of $z^{-1}$ at infinity
(see, e.g., \cite{BKA}). This equation also implies
existence of the $\tau$-function, characterized by the relation
\be
\wt S(z,\mathbf{t})=-D(z)F(\mathbf{t}),
\label{difftauKP}
\ee
and provides the dispersionless addition formula.

For the dispersionless 2DTL (d2DTL) hierarchy the $\dbar$-data are localized
on the domain $G$
which is  an annulus
$a<|z|<b$, where $a,b$ ($a,b\in\mathbb{R}$, $a,b>0$; $b>a$) 
are arbitrary,
$S_0$ 
can be represented
as \cite{TT,TT1,BKA}
\be
S_0(t,\mathbf{x},\mathbf{y})=t\log z +\sum_{n=1}^{\infty}z^n x_n+
\sum_{n=1}^{\infty}z^{-n}y_n.
\label{paramToda}
\ee
We assume that $\wt S(z)\sim \sum_{n=1}^\infty \frac{S_n}{z^n}$
as $z\rightarrow\infty$ and denote $\wt S(0)=\phi$,
$G_+=\{z,|z|>b\}$, $G_-=\{z,|z|<a\}$.
The functions $p_+=\frac{\partial S}{\partial x_1}$
and $p_-=\frac{\partial S}{\partial y_1}$ have pole singularities
while $p=\frac{\partial S}{\partial t}$ has a logarithmic singularity.
Relations, characterizing the $\tau$-function $F$, are 
$\phi=DF$, $\wt S(z_1)=-D_+(z_1)F$ $(z_1\in G_+)$,
$\wt S(z_2)=\phi-D_-(z_2)F$ $(z_2\in G_-)$, where
$D_+(z)=\sum_{n=1}^\infty\frac{1}{n}\frac{1}{z^n}\frac{\partial}{\partial x_n}$,
$D_-(z)=\sum_{n=1}^\infty\frac{1}{n}{z^n}\frac{\partial}{\partial y_n}$,
$D=\frac{\partial}{\partial t}$.
Hamilton-Jacobi equations of the hierarchy are generated by the relations
\cite{TT,TT1,BKA}
\bea
&&
e^{p(z)}-e^{p(z_1)}+z_1e^{-D_+(z_1)S(z)}=0,\quad z\in\mathbb{C},
z_1\in G_+,
\label{HJToda1}\\
&&
e^{p(z)}\left(1-e^{-D_-(z_2)S(z)}\right)=
z_2e^{(D-D_-(z_2))\phi},\quad z\in\mathbb{C},z_2\in G_-,
\label{HJToda2}
\eea
these relations also provide dispersionless addition formulae
for $F$.
%%%%%%%%%%%%%%%%%%%%%%%%%%%%%%%%%%%%%%%%%%%%%%%%%%%%%%%%%%%%%%%%%%%%
\section{$\tau$-function and symmetries}
It was noted in \cite{KM2} that equation (\ref{ddbar})
is a Lagrangian one. It can be obtained by variation of the action
(for the boundary problem in $G$)
\bea
f
=-\frac{1}{2\pi\text{i}}\iint_{G} 
\left(\frac{1}{2}\wt S_{\bar z} \wt S_z -
W_{-1}(z,\bar z,S_z)\right)dz\wedge d\bar z,
\label{action}
\eea
where
\bea
%&&
W_{-1}(z,\bar z,S_z)=\sum_{p=0}^{\infty}w_p(z,\bar z)
\frac{(S_z)^{p+1}}{p+1},
%\nn\\
%&&
\qquad
\partial_\eta W_{-1}(z,\bar z,\eta)=W(z,\bar z,\eta).
\nn
\eea
One should consider independent variations of $\wt S$,
possessing required analytic properties (analytic outside
$G$, decreasing at infinity),
keeping $S_0$ fixed.

In \cite{dtau} it was proved for dispersionless KP and
2DTL hierarchies that the action (\ref{action}) evaluated on the 
solution of nonlinear Beltrami equation (\ref{ddbar}) gives
a $\tau$-function of the corresponding hierarchy. Here we formulate
first an auxiliary statement, valid for 
an hierarchy with arbitrary $G$, and then
recall the results of \cite{dtau}.
\begin{lemma}
Let us evaluate the action (\ref{action}) on the solution of
the boundary problem for (\ref{ddbar}) and consider it as a functional 
of $S_0$,
\bea
F(S_0)
=-\frac{1}{2\pi\mathrm{i}}\iint_{G} 
\left(\frac{1}{2}\wt S_{\bar z} \wt S_z-
W_{-1}(z,\bar z,S_z)\right)dz\wedge d\bar z.
\label{TAU0}
\eea
Then
\be
\delta F=-\frac{1}{2\pi\mathrm{i}}\int_{\partial G}
\wt S\; d \delta S_0.
\label{tauvar}
\ee
\end{lemma}
Using the formula (\ref{tauvar}), it is easy to reproduce
the statements of the work \cite{dtau}.
%%%%%%%%%%%%%%%%%%%%%%%%%%%%%%%%%%%%%%%%%%%%%%%%%%%%%%%%%%%%%%%%%%%%%%%
\begin{theorem}
The function
\bea
F(\mathbf{t})
=-\frac{1}{2\pi\mathrm{i}}\iint_{D} 
\left(\frac{1}{2}\wt S_{\bar z}(\mathbf{t}) \wt S_z(\mathbf{t})-
W_{-1}(z,\bar z,S_z(\mathbf{t}))\right)dz\wedge d\bar z,
\label{TAU}
\eea
i.e., the action (\ref{action}) evaluated on the solution 
of the problem (\ref{ddbar}), 
with $S_0$ given by (\ref{paramKP}), is a $\tau$-function of 
dKP hierarchy.
\end{theorem}
%%%%%%%%%%%%%%%%%%%%%%%%%%%%%%%%%%%%%%%%%%%%%%%%%%%%%%%%%%%%%%%%%%%%%%
\begin{theorem}
The function
\bea
F(t,\mathbf{x},\mathbf{y})
=\frac{-1}{2\pi\mathrm{i}}\iint_{G} 
\left(\frac{1}{2}\wt S_{\bar z}(t,\mathbf{x},\mathbf{y}) 
\wt S_z(t,\mathbf{x},\mathbf{y})-
W_{-1}(z,\bar z,S_z(t,\mathbf{x},\mathbf{y}))\right)dz\wedge d\bar z,
%\label{TodaTAU}
\nn
\eea
with $S_0$ given by (\ref{paramToda}),
evaluated on the solution 
of the problem (\ref{ddbar}), 
is a $\tau$-function of dispersionless 2DTL hierarchy
($G$ is an annulus defined above).
\end{theorem}
%%%%%%%%%%%%%%%%%%%%%%%%%%%%%%%%%%%%%%%%%%%%%%%%%%%%%%%%%%%%%%%%%%%%%%%
The function $W$ is the $\dbar$ data for the quasiclassical
$\dbar$-problem.
Its variations provide us with a wide class of variations of the
function $F$ (first we will use formula (\ref{TAU0}) defined for
arbitrary $G$).

For the functions $W$ of the form (\ref{Wform}),
varying $w_n(z,\bar z)$,
one has 
\beaa
\delta W=\sum_{n=1}^{\infty}\delta w_n(z,\bar z)\left(S_z\right)^n,
%\\
\qquad
\delta W_{-1}=\sum_{n=1}^{\infty}
\frac{\delta w_n}{n+1}\left(S_z\right)^{n+1},
\eeaa
and 
\bea
\delta F
%=\frac{1}{2\pi\text{i}}\iint_{G}
%\sum_{p=0}^{\infty}
%\delta w_p(z,\bar z)\frac{S_z^{p+1}}{p+1}dz\wedge d\bar z
=\frac{\epsilon}{2\pi\text{i}}
\iint_{G} U(z,\bar z,S_z)dz\wedge d\bar z,
\label{Fvar}
\eea
where $U(z,\bar z,\eta)$ is an \textit{arbitrary} function
with the support in $z$-plane belonging to $G$,
analytic in $\eta$. The
formula (\ref{Fvar}) gives a general variation (infinitesimal
symmetry transformation) of the $\tau$-function. Let us consider
some special symmetry transformations.

Considering elementary variation 
$\delta w_{n_0}=\epsilon \alpha_{n_0} \delta(z-z_0)$,
$\delta w_n=0, n\neq n_0$, one gets
\bea
\delta F=\frac{\epsilon}{2\pi\text{i}}\frac{\alpha_{n_0}}{(n_0+1)}
 \left(S_z\right)^{n_0+1}|_{z=z_0},
\label{Fvar1}
\eea
and, respectively,
\bea
\delta \wt S=-\frac{\epsilon}{2\pi\text{i}}\frac{\alpha_{n_0}}{(n_0+1)}
D(z)\left(S_z(z_0)\right)^{n_0+1}.
\label{Svar}
\eea
Taking superposition of elementary variations (\ref{Fvar1}),
we obtain a variation of the form
\bea
\delta F=\epsilon f(S_z(z_0)),
\label{Fvar2}
\eea
where $f$ is an arbitrary analytic function (summation over
different points and integration over $z_0$ are also possible).

Another simple symmetry transformation is given by
\bea
\delta F=\epsilon c(S(z_2)-S(z_1)),
\label{Fvar3}
\eea
where $z_1,z_2$ are arbitrary points belonging to the domain $G$, 
and $c$ is some constant. Indeed,
\beaa
&&
S(z_2)-S(z_1)=\int^{z_2}_{z_1}d(S)
=
\int^{z_2}_{z_1}S_z dz + \int^{z_2}_{z_1}S_{\bar z} d{\bar z}\\
&&\qquad
=
\int^{z_2}_{z_1}S_z dz + \int^{z_2}_{z_1}W(z,\bar z,S_z) d{\bar z},
\eeaa
thus the function $S(z_2)-S(z_1)$ corresponds to the special case
of general variation (\ref{Fvar}). In the limit $z_2\rightarrow z_1$
the transformation (\ref{Fvar3}) is reduced to elementary symmetries
(\ref{Fvar1}).
%%%%%%%%%%%%%%%%%%%%%%%%%%%%%%%%%%%%%%%%%%%%%%%%%%%%%%%%%%%%%%%%%%%%%%%
\section{Symmetry constraints}
Given infinitesimal symmetries of the form
$$
\delta F=\epsilon \Phi,
$$
(cf., (\ref{Fvar}), (\ref{Fvar2}), (\ref{Fvar3})),
it is possible to introduce symmetry constraints
\be
\frac{\partial}{\partial t_i}F=\Phi.
\label{constraint}
\ee
These constraints are preserved by the flows of the hierarchy
and represent a reduction of the hierarchy. There is a lot
of similarity between constraints (\ref{constraint})
and symmetry constraints
in the (2+1)-dimensional dispersionfull case, which are rather well studied
\cite{KStr}-\cite{Orlov91}. Below we will demonstrate that constraints
(\ref{constraint}) lead to hydrodynamic 
reductions, and give a regular way to introduce basic 
objects connected with these reductions.

Technically, in the dispersionful case one starts with
linear equations of the hierarchy (which give the hierarchy itself in the form of compatibility conditions). Symmetry constraints connect
the potentials in these equations with some special wave functions,
and in the end one obtains (1+1)-dimensional integrable system
for the set of wave functions. The simplest symmetry constraint for
KP hierarchy leads to nonlinear Schr\"odinger equation.

In the dispersionless case we will follow the same strategy.
The role of linear equations of the hierarchy is
played by Hamilton-Jacobi equations of the dispersionless hierarchy.
%%%%%%%%%%%%%%%%%%%%%%%%%%%%%%%%%%%%%%%%%%%%%%%%%%%%%%%%%%%%%%%%%%%
\section{Zakharov reduction for the dKP hierarchy}
%%%%%%%%%%%%%%%%%%%%%%%%%%%%%%%%%%%%%%%%%%%%%%%%%%%%%%%%%%%%%%%%%%%%
We begin with a simple example of symmetry constraint for the
dKP hierarchy.
The first Hamilton-Jacobi equation of dKP hierarchy
(dispersionless analogue of Lax equation) is
\be
S_y=p^2+2u; \quad p=S_x,\; u=-\partial_x \wt S_1,
\label{HJ1}
\ee
where $x=t_1$, $y=t_2$.
Let us consider the symmetry constraint
\be
F_x=f(S_z(z_0)),
\label{sc1}
\ee
which is equivalent to the condition (cf., (\ref{difftauKP}))
$$
u=\partial_x f(S_z(z_0)).
$$
Differentiating (\ref{HJ1}) with respect to $z$ and evaluating the result
at $z=z_0$,
we get 
$$
\frac{\partial_y S_z(z_0)}{\partial_x S_z(z_0)}=2p_0,\quad p_0=p(z_0).
$$
Using this equation and evaluating (\ref{HJ1}) at $z=z_0$,
we obtain a system of hydrodynamic type
\bea
\left\{
\begin{array}{l}
\partial_y p_0=\partial_x (p_0^2+2u),
\\
\partial_y u=2\partial_x (p_0 u).
\end{array}
\right.
\label{hydro1}
\eea
This system is well known \cite{Z81}, it represents a dispersionless
limit of nonlinear Schr\"odinger eqution. Thus a constraint
(\ref{sc1}) represents a proper dispersionless analogue of the
simplest symmetry constraint of KP hierarchy.

It is possible to move further and demonstrate that the symmetry constraint (\ref{sc1}) leads to explicit expression for
the function $z(p)$, which plays a fundamental role in the picture
of constrained hierarchy and allows to find Riemann invariants
and hodograph equations. To do this, we will use generating
Hamilton-Jacobi equation (\ref{DE2}) rewritten as
\be
D(z)S(z')=-\log\frac{p(z)-p(z')}{z},\quad |z|>1,
\label{HJgenKP}
\ee
which produces infinite set of Hamilton-Jacobi equations
by expansion into the powers of $z^{-1}$ at infinity
(see, e.g., \cite{BKA}). Differentiating this equation by
$z'$ and evaluating the result at $z'=z_0$, we get
$$
\frac{D(z)S_z(z_0)}{\partial_x S_z(z_0)}=\frac{1}{p-p_0}.
$$
Then, using the constraint (\ref{sc1}) and simple formula
$$
D(z)F_x=z-p,
$$
we obtain the relation
\be
z=p+\frac{u}{p-p_0}.
\label{zKP}
\ee
The function $z(p)$ (Sato function) 
plays a crucial role in the picture of dispersionless
hierarchy. Dispersionless KP hierarchy (for general $z(p)$)
can be written in the form 
$
\partial_n z(p)=\{z^n_+,z\}.
$
Relation (\ref{zKP}) defines the function $z(p)$ for
the constrained hierarchy,
it is well known \cite{Z81} and is usually called Zakharov reduction.
Thus we have demonstrated that symmetry constraint (\ref{sc1})
leads to Zakharov reduction (\ref{zKP}).

Riemann invariants for the constrained hierarchy of equations
of hydrodynamic type (the first of which is (\ref{hydro1})),
are given by the values of $z(p)$ (\ref{zKP}) at $p_i$ where
$\frac{\partial z}{\partial p}$ vanishes,
$\lambda_i=z(p_i)$, $\frac{\partial z}{\partial p}(p_i)=0$.
It is rather straightforward to introduce hodograph
transform for the solution of constrained hierarchy
\cite{14}.
%%%%%%%%%%%%%%%%%%%%%%%%%%%%%%%%%%%%%%%%%%%%%%%%%%%%%%%%%%%
\section{Generic symmetry constraint for the dKP hierarchy}
%%%%%%%%%%%%%%%%%%%%%%%%%%%%%%%%%%%%%%%%%%%%%%%%%%%%%%%%%%%%%%%%%
Now we will use the symmetry (\ref{Fvar3}) and consider the constraint
\be
F_x=\sum_{i=1}^N c_i (S_i-\tilde S_i),
\label{sc2}
\ee
where $S_i=S(z_i)$, $\tilde S_i=S(\tilde z_i)$, $z_i$, $\tilde z_i$ are some sets of
points, and $c_i$ are arbitrary constants.  In equivalent form,
$$
u=\partial_x\sum_{i=1}^N c_i (S_i-\tilde S_i)=
\sum_{i=1}^N c_i (p_i-\tilde p_i).
$$
Evaluating Hamilton-Jacobi equation (\ref{HJ1}) 
and its higher counterpart
$$
S_t=p^3 + 3up - \frac{3}{2}\wt S_1,
$$
where $t=t_3$,
at $z$ equal to
$z_i$, $\tilde z_i$, we immediately get two systems of hydrodynamic type
\be
\left\{
\begin{array}{l}
\displaystyle
\partial_y p_k=\partial_x
\bigl((p_k^2)+2\sum_{i} c_i (p_i-\tilde p_i)\bigr)
\\ 
\displaystyle
\partial_y \tilde p_k=\partial_x
\bigl(({\tilde p}_k^2)+2\sum_{i} c_i (p_i-{\tilde p}_i)\bigr)
\end{array}
\right.
\quad k=1,\dots, N
\label{hydro2}
\ee
and 
\be
\left\{
\begin{array}{l}
\displaystyle
\partial_t p_k=\partial_x
\bigl((p_k^3)+3p_k\sum_{i} c_i (p_i-\tilde p_i)
+\frac{3}{2}\sum_{i}c_i(p_i^2-\tilde p_i^2) \bigr)
\\ 
\displaystyle
\partial_t \tilde p_k=\partial_x
\bigl((\tilde p_k^3)+3\tilde p_k\sum_{i} c_i (p_i-\tilde p_i)
+\frac{3}{2}\sum_{i}c_i(p_i^2-\tilde p_i^2) \bigr)
\end{array}
\right.
\label{hydro2a}
\ee
These systems can be written in Hamiltonian form,
\bea
&&
\partial_{t_n} \mathbf{p}=J \frac{\delta H_{n+1}}{\delta \mathbf{p}},
\quad
J=
\begin{pmatrix}
~C^{-1} & 0\\
0 & -C^{-1}
\end{pmatrix}\partial_x,
\nn\\&&
H_3=\frac{1}{3}\int dx\left(\sum_ic_i(p_i^3-{\tilde p}_i^3)
+ 3\bigl(\sum_{i} c_i (p_i-{\tilde p}_i)\bigr)^2\right),
\nn\\
&&
H_4=\frac{1}{4}\int dx\left(\sum_ic_i(p_i^4-{\tilde p}_i^4)
+  6\sum_i c_i(p_i-\tilde p_i)\times\sum_i c_i (p_i^2-{\tilde p}_i^2)
\right),
\nn
\eea
where 
$\mathbf{p}=(p_1,\dots,p_N,\tilde p_1,\dots,\tilde p_N)^\mathrm{t}$,
$C$ is diagonal $N\times N$ matrix with the entries $c_i$,
$C_{ik}=c_i\delta_{ik}$, $i,k=1,\dots, N$. 

Common solution of the systems (\ref{hydro2}), (\ref{hydro2a})
generates a solution $u=\sum_i c_i(p_i-\tilde p_i)$
of the dKP equation.
%%%%%%%%%%%%%%%%%%%%%%%%%%%%%%%%%%%%%%%%%%%%%%%%%%%%%%%%%%%%%%%%%%%%

The Sato function $z(p)$, which allows to construct Riemann invariants
and higher Hamiltonians, is the central 
object for the constrained hierarchy. 
To find it, we, similar to the first
example, use the generating Hamilton-Jacobi equation (\ref{HJgenKP}).
Evaluating this equation at $z'$ equal to $z_i$, $\tilde z_i$
and combining the results, we get
\be
D(z)\sum_{i=1}^N c_i (S_i-\tilde S_i)=-\sum_{i=1}^N c_i
\log\frac{p-p_i}{p-\tilde p_i}.
\ee
Then, due to the constraint (\ref{sc2}),
$$
D(z)F_x=-\sum_{i=1}^N c_i
\log\frac{p-p_i}{p-\tilde p_i},
$$
and finally we obtain the function $z(p)$,
\be
z=p-\sum_{i=1}^N c_i
\log\frac{p-p_i}{p-\tilde p_i}.
\label{zKP2}
\ee
Equation $\frac{\partial z}{\partial p}=0$ is an algebraic one,
having $2N$ roots $p_\alpha$, and Riemann invariants 
for the constrained hierarchy are $\lambda_\alpha=z(p_\alpha)$.

Expansion of $z(p)$ at $p\rightarrow\infty$ looks like 
\be
z=p+\sum_{n=1}^{\infty}\frac {v_n}{p^{n}},
\quad v_n=\sum_{i=1}^{N}\frac{c_i(p_i^n- \tilde p_i^n)}{n}.
\label{zKP2a}
\ee

Equation (\ref{HJgenKP}) allows us to obtain the generating
equation for the whole hierarchy of systems of hydrodynamic
type associated with the constraint (\ref{sc2}). Indeed,
differentiating (\ref{HJgenKP}) with respect to $x$ and evaluating
it at the points $z_i$, $\tilde z_i$, one gets
\be
\left\{
\begin{array}{l}
\displaystyle
D(z) p_k=-\partial_x
\log(p-p_k)
\\ 
\displaystyle
D(z) \tilde p_k=-\partial_x
\log(p-\tilde p_k)
\end{array}
\right.
\quad k=1,\dots, N,
\label{hydro2gen}
\ee
where $p$ is a function of $z$,
$(p_1,\dots,p_N,\tilde p_1,\dots,\tilde p_N)$,
defined by the relation (\ref{zKP2}). Expanding both sides
of this system into the powers of ${z}^{-1}$, one gets the systems (\ref{hydro2}),
(\ref{hydro2a}) and their higher counterparts. Introducing
to $S_0(z,\mathbf{t})$ dependence on extra time $t_{\log}$,
$S_0(z')\rightarrow S_0(z') - t_{\log} \log(1-\frac{z'}{z})$,
$\frac{\partial}{\partial t_{\log}}=D(z)$, we may consider
(\ref{hydro2gen}) as a system of hydrodynamic type with the
time $t_{\log}$. Hamiltonian for this system is given by the
expression
$$
H_{\log}=\int dx\left(\frac{1}{2}p^2-\sum_{i=1}^N c_i 
\bigl(\tilde p_i-p_i
+p_i\log(p-p_i)-\tilde p_i\log(p-\tilde p_i)
\bigr)\right).
$$ 

Differentiating (\ref{hydro2gen}) by $z$, we will get
another form of generating system,
\be
\left\{
\begin{array}{l}
\displaystyle
D_z(z) p_k=-\partial_x
\frac{p_z}{p-p_k}
\\ 
\displaystyle
D_z(z) \tilde p_k=-\partial_x
\frac{p_z}{p-\tilde p_k}
\end{array}
\right.
\quad k=1,\dots, N,
\label{hydro2gena}
\ee
where $p,p_z$ are functions of $z$,
$(p_1,\dots,p_N,\tilde p_1,\dots,\tilde p_N)$,
defined by the relation (\ref{zKP2}).
This system can be considered as a system of hydrodynamic type with the time
$t_\text{pole}$, $S_0(z')\rightarrow S_0(z')+t_\text{pole}
(z-z')^{-1}$, $\frac{\partial}{\partial t_\text{pole}}=D_z(z)$.
Hamiltonian for the generating system (\ref{hydro2gena})
is remarkably simple,
$$
H_\text{pole}=\int p\,dx ,
$$
and its expansion into the powers of $z^{-1}$ gives a general
formula for the Hamiltonians of constrained hierarchy,
\be
H_n=\frac{1}{n}\int \mathrm{res}_{\infty}\bigl(z(p)^n\bigr)\; dx =
\partial_x \wt S_n,
\label{Hgen}
\ee
which is in agreement with general results \cite{KM77}.

In a similar way one can treat the symmetry constraints
\be
\frac{\partial F}{\partial t_n}=\sum_{i=1}^N c_i (S_i-\tilde S_i).
\label{sc2n}
\ee
In particular, since
$$
\frac{\partial S}{\partial t_n}=p^n+u_{n-2}p^{n-2}+\dots+u_0,
$$
one obtains a corresponding Sato function
\be
E(p)=z^n=p^n+u_{n-2}p^{n-2}+\dots+u_0  -\sum_{i=1}^N c_i
\log\frac{p-p_i}{p-\tilde p_i}.
\label{SatoKPn}
\ee
\begin{small}
\noindent
\textbf{Remark.} It is easy to check that in the limit
$\tilde z_i=z_i+\epsilon_i$ $\epsilon_i\rightarrow 0$, 
$c_i\epsilon_i=\text{const}$, the Sato function (\ref{SatoKPn})
reproduces the curve
\be
z^n=p^n+u_{n-2}p^{n-2}+\dots+u_0 +\sum_{i=1}^N 
\frac{a_i}{p-p_i},
\label{Zn}
\ee
which has been discussed in \cite{14,16}. At $n=1$
one gets a general Zakharov reduction.
\end{small}
%%%%%%%%%%%%%%%%%%%%%%%%%%%%%%%%%%%%%%%%%%%%%%%%%%%%%%%%%%%%%%%%
\section{Symmetry constraint for the d2DTL hierarchy}
%%%%%%%%%%%%%%%%%%%%%%%%%%%%%%%%%%%%%%%%%%%%%%%%%%%%%%%%%%%%%%%
Now we will consider symmetry reduction of d2DTL hierarchy 
defined by the constraint
\be
F_t=\sum_{i=1}^N c_i (S_i-\tilde S_i).
\label{sc3}
\ee
Technically this case is very similar to that considered
in the previous section,
and we will omit some details.
The first Hamilton-Jacobi equations of d2DTL hierarchy are
\bea
&&
\partial_x S(z)=
e^{p}-
\partial_t \wt S_1,
\label{HJ01}
\\&&
\partial_y S=e^{\phi_t-
p},
\label{HJ02}
\eea
where $x=x_1$, $y=y_1$. Using the constraint (\ref{sc3}),
from these equations we get two systems of hydrodynamic type,
namely
\bea
\left\{
\begin{array}{l}
\partial_x p_i=\partial_t\bigl(e^{p_i}+
\sum_k c_k(e^{p_k}-e^{\tilde p_k})\bigr)
\vspace{1\jot}\\
\partial_x \tilde p_i=\partial_t\bigl(e^{\tilde p_i}+
\sum_k c_k(e^{p_k}-e^{\tilde p_k})\bigr)
\end{array}
\right.
\qquad i=1,\dots,N
\label{hydro3x}
\eea
and
\bea
\left\{
\begin{array}{l}
\partial_y p_i=\partial_t 
\exp\bigl(-p_i + \sum_k c_k(p_k-\tilde p_k)\bigr)
\vspace{1\jot}\\
\partial_y \tilde p_i=\partial_t 
\exp\bigl(-\tilde p_i+\sum_k c_k(p_k-\tilde p_k)\bigr)
\end{array}
\right.
\qquad i=1,\dots,N
\label{hydro3y}
\eea
Both systems can be written in Hamiltonian form,
\bea
&&
\partial_x \mathbf{p}=J \frac{\delta H_1^+}{\delta \mathbf{p}},
\quad
\partial_y \mathbf{p}=J \frac{\delta H_1^-}{\delta \mathbf{p}},
%\nn\\&&
\quad
J=
\left( E+\begin{pmatrix}
~C^{-1} & 0\\
0 & -C^{-1}
\end{pmatrix}\right)\partial_t,
\nn
\eea
where $E$ is $2N\times2N$ matrix with all entries
equal to $1$, $E_{ik}=1$, $i,k=1,\dots, 2N$,
$C$ is diagonal $N\times N$ matrix with the entries $c_i$,
$C_{ik}=c_i\delta_{ik}$, $i,k=1,\dots, N$,
and the Hamiltonians are, respectively,
\bea
&&
H_1^+=\int dt \sum_k c_k(e^{p_k}-e^{\tilde p_k}),
\label{H1}
\\
&&
H_1^-=\int dt\, \exp\left(\sum_k c_k(p_k-\tilde p_k)\right)
\sum_k c_k(e^{- \tilde p_k}-e^{- p_k}).
\label{H2}
\eea

Using generating equation (\ref{HJToda1}), we obtain
the function $z(p)$ for the constrained hierarchy,
\be
z=e^p\prod_{k=1}^{N}
\left(\frac{e^p-e^{p_k}}{e^p-e^{\tilde p_k}}\right)^{-c_k}.
\ee
Starting with second generating equation (\ref{HJToda2}),
we get equivalent form of this relation,
\be
z=\exp\bigl(p-\sum_{k=1}^N c_k(p_k-\tilde p_k)\bigr)\prod_{k=1}^{N}
\left(\frac{e^{-p}-e^{-p_k}}{e^{-p}-e^{-\tilde p_k}}\right)^{-c_k}.
\ee
Equation $\frac{\partial z}{\partial p}=0$ is algebraic (with respect to the variable $\xi=e^p$),
having $2N$ roots $\xi_\alpha$, and Riemann invariants 
for the constrained hierarchy are $\lambda_\alpha=z(\xi_\alpha)$.

Generating systems for the constrained hierarchy
read
\be
\left\{
\begin{array}{l}
\displaystyle
D_+(z) p_k=-\partial_t
\log(e^p-e^{p_k})
\\ 
\displaystyle
D_+(z) \tilde p_k=-\partial_t
\log(e^p-e^{\tilde p_k})
\end{array}
\right.
\quad k=1,\dots, N,
\label{hydro2genT1}
\ee
\be
\left\{
\begin{array}{l}
\displaystyle
D_-(z) p_k=-\partial_t
\log(1-e^{p-p_k})
\\ 
\displaystyle
D_-(z) \tilde p_k=-\partial_t
\log(1-e^{p-\tilde p_k})
\end{array}
\right.
\quad k=1,\dots, N,
\label{hydro2genT2}
\ee
(`logarithmic' form)
and
\be
\left\{
\begin{array}{l}
\displaystyle
D_{z\pm}(z) p_k=\partial_t
\frac{p_z}{e^{p_k-p}-1}
\\ 
\displaystyle
D_{z\pm}(z) \tilde p_k=\partial_t
\frac{p_z}{e^{\tilde p_k-p}-1}
\end{array}
\right.
\quad k=1,\dots, N,
\label{hydro2genT1a}
\ee
(`pole' form). Hamiltonian for the `pole' form
of generating equations is
$$
H_\pm(z)=\frac{1}{z}\int p\, dt.
$$
Expansion of this Hamiltonian at $z=\infty$ and $z=0$ gives
general expressions for the Hamiltonians of constrained hierarchy, 
\bea
&&
H_n^+=\frac{1}{n}\int \mathrm{res}_{\infty}
\bigl(z(\xi)^n\xi^{-1}\bigr)\;dt
\label{Hgen1},\\
&&
H_n^-=\frac{1}{n}\int  \mathrm{res}_{0}
\bigl(z(\xi)^{-n}\xi^{-1}\bigr)\;dt,
\label{Hgen2}
\eea
where $\xi=e^p$.

In more detail, the Hamiltonian and bi-Hamiltonian structure 
of equation considered above will be considered elsewhere.

Symmetry constraints of the type $F_{x_n}=\sum_{i=1}^N c_i (S_i-\tilde S_i)$ and $F_{y_n}=\sum_{i=1}^N c_i (S_i-\tilde S_i)$ are treated
analogously. The corresponding Sato function is given by 
(\ref{SatoKPn}) with substitutions $p\rightarrow e^p$,
$p\rightarrow e^{-p}$, respectively.
%%%%%%%%%%%%%%%%%%%%%%%%%%%%%%%%%%%%%%%%%%%%%%%%%%%%%%%%%%%%%%%%%%%%%%%%
\section{Conclusion}
As we have seen, the systems of hydrodynamic type and associated
complex curves (Sato functions) are direct consequences of
the generating Hamilton-Jacobi equations (\ref{DE2}),
(\ref{HJToda1},\ref{HJToda2}) under the symmetry constraint.

We would like to note here that complex curves discussed above
are connected with the $\dbar$-equation in a very natural way.
Let us consider the dKP hierarchy as an example. The function 
$p=\partial_x S$ solves a linear Beltrami equation
$p_{\bar z}=W'p_z$, and under certain conditions is a basic
homeomorphism. The inverse function $z(p,\bar p)$ is analytic in some
neighborhood of infinity. The Cauchy-Green formula implies that
\be
z=p+\frac{1}{2\pi\mathrm{i}}\iint 
\frac{d p'\wedge d \bar p'}{p'-p}\frac{\partial z}{\partial \bar p'}.
\label{int}
\ee
In the neighborhood of infinity one has
\be
z=p+\sum_{n=1}^{\infty}\frac{v_n}{p^n},
\ee
where
\be
v_n=-\frac{1}{2\pi\mathrm{i}}\iint {p'}^{n-1}\frac{\partial z}{\partial \bar p'}d p'\wedge d \bar p'.
\label{int2}
\ee
Let us choose a special $\frac{\partial z}{\partial \bar p}$
with a support along the curve $\Gamma=\bigcup_i \Gamma_i$
composed of $N$ disconnected pieces $\Gamma_i$ with the ends at the 
points
$z_i$ and $\tilde z_i$ and such that 
$\frac{\partial z}{\partial \bar p}=-\sum_{i=1}^N c_i \delta_{\Gamma_i}$.
For such `constrained' $\dbar$-data formula (\ref{int})
gives
$$
z=p-\sum_{i=1}^N c_i
\log\frac{p-p_i}{p-\tilde p_i},
%\label{zKP2}
$$
while from (\ref{int2}) we obtain
$$
\quad v_n=\sum_{i=1}^{N}\frac{c_i(p_i^n- \tilde p_i^n)}{n},
$$
that coincides with (\ref{zKP2},\ref{zKP2a}). For the $\dbar$-data
given by $\frac{\partial z}{\partial \bar p}=-\sum_{i=1}^N
a_i\delta(p-p_i)$, equation (\ref{int}) immediately leads to the
Sato function associated with Zakharov reduction (see (\ref{Zn})).
Thus, the Cauchy-Green formula is a generator of complex curves
associated with equations of hydrodynamic type.

Equation (\ref{int}) is the two-dimensional extension
of the integral equation discussed in \cite{Z81}, \cite{Gibb81}.
%%%%%%%%%%%%%%%%%%%%%%%%%%%%%%%%%%%%%%%%%%%%%%%%%%%%%%%%%%%%%%%%%%%%%
\subsection*{Acknowledgments}
Authors would like to thank Luis Mart\'\i nez Alonso
for useful discussions. LVB is also grateful to E.V.
Ferapontov for attracting his interest to the problem of
hydrodynamic reductions.
LVB was supported in part by RFBR grant
01-01-00929 and President of Russia grant 1716-2003; BGK was supported in part
by the grant COFIN 2002 `Sintesi'. 
%%%%%%%%%%%%%%%%%%%%%%%%%%%%%%%%%%%%%%%%%%%%%%%%%%%%%%%%%%%%%%%%%%%%%%%%%%%%%%%

\end{document}